\documentclass{llncs}
\usepackage{llncsdoc}

\input{tcilatex}
\begin{document}

\title{A bird's eye view of \\
Matrix Distributed Processing}
\author{Massimo Di Pierro}
\institute{{\footnotesize School of {\bf C}omputer Science, {\bf T}elecommunications and {\bf I}nformation Systems}\\
{\footnotesize DePaul University, 243 S. Wabash Av., Chicago, IL 60604, USA}}
\maketitle

\begin{abstract}
We present Matrix Distributed Processing, a C++ library for fast development
of efficient parallel algorithms. {\tt MDP} is based on MPI and consists of a
collection of C++ classes and functions such as lattice, site and field.
Once an algorithm is written using these components the algorithm is
automatically parallel and no explicit call to communication functions is
required. {\tt MDP} is particularly suitable for implementing parallel solvers for
multi-dimensional differential equations and mesh-like problems.
\end{abstract}

\section{Introduction}

Matrix Distributed Processing ({\tt MDP}) \cite{mdp1} is a collection of classes
and functions written in C++ to be used as components for fast development
of efficient parallel algorithms. Typical algorithms include solvers for
partial differential equations, mesh-like algorithms and various types of
graph-based problems. These algorithms find frequent application in many
sectors of physics, engineering, electronics and computational finance.

{\tt MDP} components can be divided into two main categories:

\begin{itemize}
\item[$\bullet$]  Non parallel components: Linear Algebra components (class \texttt{%
mdp\_complex}, class \texttt{mdp\_array}, class \texttt{mdp\_matrix}) and
Statistical Analysis components (class \texttt{Measure}, class \texttt{%
Jackboot})

\item[$\bullet$]  Parallel components: (class \texttt{mdp\_lattice}, class \texttt{%
mdp\_site}, class \texttt{mdp \_field}, etc.)
\end{itemize}

In this paper we will focus exclusively on the Linear Algebra and the
Parallel components\footnote{%
The parallel components can interoperate with other third party C/C++ linear
algebra packages and can be used to parallelize existing applications with
minimal effort.}.

{\tt MDP} is based on MPI and can be used on any machine with an ANSI C++ and
support for the MPI communication protocol. No specific communication
hardware is required but a fast network switch is suggested. {\tt MDP} has been
tested on Linux PC clusters, SUN workstations and a Cray T3E.

The best way to introduce {\tt MDP} is to write a program that solves a typical
problem:

\textbf{Problem:} Let's consider the following differential equation: 
\begin{equation}
\nabla ^2\varphi (x)=f(x)  \label{eq1}
\end{equation}
where $\varphi (x)$ is a field of $2\times 2$ Complex matrices defined on a
3D space (\texttt{space}), $x=(x_0,x_1,x_2)$ limited by $0\leq x_i<L_i$, and 
\begin{eqnarray}
L &=&\{10,10,10\}, \\
f(x) &=&A\sin (2\pi x_1/L_1),  \nonumber \\
A &=&\left( 
\begin{array}{ll}
1 & i \\ 
3 & 1
\end{array}
\right)  \nonumber
\end{eqnarray}
The initial conditions are $\varphi _{initial}(x)=0.$ We will also assume
that $x_i+L_i=x_i$ (torus topology).

\textbf{Solution:} In order to solve eq.~(\ref{eq1}) we first discretize the
Laplacian ($\nabla ^2=\partial _0^2+\partial _1^2+\partial _2^2$) and
rewrite it as 
\begin{equation}
\sum_{\mu =0,1,2}\left[ \varphi (x+\widehat{\mu })-2\varphi (x)+\varphi (x-%
\widehat{\mu })\right] =f(x)
\end{equation}

where $\widehat{\mu }$ is a unit vector in the discretized space in
direction $\mu $. Hence we solve it in $\varphi (x)$ and obtain the
following a recurrence relation 
\begin{equation}
\varphi (x)=\frac{\sum_{\mu =0,1,2}\left[ \varphi (x+\widehat{\mu })+\varphi
(x-\widehat{\mu })\right] -f(x)}6  \label{eq3}
\end{equation}

The following is a typical {\tt MDP} program that solves eq.~(\ref{eq1}) by recursively iterating eq.~(\ref{eq3}). The program is parallel but there are no explicit call to communication functions:
\begin{verbatim}
00    #include "mdp.h"              
01
02    void main(int argc, char** argv) {
03       mdp.open_wormholes(argc,argv);    // open communications
04       int L[]={10,10,10};               // declare volume
05       mdp_lattice      space(3,L);      // declare lattice
06       mdp_site         x(space);        // declare site variable
07       mdp_matrix_field phi(space,2,2);  // declare field of 2x2
08       mdp_matrix       A(2,2);          // declare matrix A
09       A(0,0)=1;  A(0,1)=I;
10       A(1,0)=3;  A(1,1)=1;
11
12       forallsites(x)                    // loop (in parallel)
13          phi(x)=0;                      // initialize the field
14       phi.update();                     // communicate!
15
16       for(int i=0; i<1000; i++) {       // iterate 1000 times
17          forallsites(x)                 // loop (in parallel)
18             phi(x)=(phi(x+0)+phi(x-0)+
19                     phi(x+1)+phi(x-1)+
20                     phi(x+2)+phi(x-2)-
21                     A*sin(2.0*Pi*x(1)/L[1]))/6;  // equation
22          phi.update();                  // communicate!
23       }
24       phi.save("field_phi.mdp");        // save field
25       mdp.close_wormholes();            // close communications
26    }
\end{verbatim}

\textbf{Notes:}

\begin{itemize}
\item[$\bullet$]  Line 00 includes the {\tt MDP} library.

\item[$\bullet$]  Lines 03 and 25 respectively open and close the communication
channels over the parallel processes.

\item[$\bullet$]  Line 04 declares the size of the box $L=\{L_0,L_1,L_2\}$

\item[$\bullet$]  Line 05 declares a lattice, called \texttt{space}, 3-dimensional, on
the box $L$. {\tt MDP} supports up to 10-dimensional lattices. By default a
lattice object is a mesh with torus topology. It is possible to specify an
alternative topology, boundary conditions and any parallel partitioning for
the lattice. Notice that each lattice object contains a parallel random
generator.

\item[$\bullet$]  Line 06 declares a variable site, called \texttt{x}, that will be
used to loop over lattice sites (in parallel).

\item[$\bullet$]  Line 07 declares a field of $2\times 2$ matrices, called \texttt{phi}%
, over the lattice \texttt{space}. {\tt MDP} is not limited to fields of matrices.
It is easy to declare fields of any user-defined structure or class.

\item[$\bullet$]  Lines 08 through 10 define the matrix A.

\item[$\bullet$]  Lines 12 and 13 initialize the field \texttt{phi}. Notice that 
\texttt{phi} is distributed over the parallel processes and \texttt{%
forallsites} is a parallel loop.

\item[$\bullet$]  Line 14 performs communications so that each process becomes aware of
changes in the field performed by other processes (\emph{synchronization}).

\item[$\bullet$]  Lines 16 through 24 perform 1000 iterations to guarantee convergence.
In real life applications one may want to implement some convergence
criteria as stopping condition.

\item[$\bullet$]  Line 17 loops over all sites in parallel.

\item[$\bullet$]  Lines 18 through 21 implement eq.~(\ref{eq3}). Notice the similarity
in notation. Here \texttt{phi(x)} is a $2\times 2$ complex matrix

\item[$\bullet$]  Line 22 performs \emph{synchronization}.

\item[$\bullet$]  Line 24 saves the field. Notice than any field, including the user
defined ones, inherit methods \texttt{save} and \texttt{load} from a basic
class \texttt{mdp\_field}.

\item[$\bullet$]  It should also be noted that all {\tt MDP} classes and functions are both
type and exception safe. Moreover {\tt MDP} components can be used without
knowledge of C pointers and pointer arithmetics.
\end{itemize}

\newpage
\section{Linear Algebra}

{\tt MDP} includes a Linear Algebra package. The basic classes are:

\begin{itemize}
\item[$\bullet$]  class \texttt{mdp\_real}, that should be use in place of float or
double.

\item[$\bullet$]  class \texttt{mdp\_complex}, (just another implementation of complex
numbers).

\item[$\bullet$]  class \texttt{mdp\_array}, for vectors and/or multidimensional
tensors.

\item[$\bullet$]  class \texttt{mdp\_matrix}, for any kind of complex rectangular
matrix.
\end{itemize}

The most notably difference between our linear algebra package and other
existing packages is its natural syntax.

For example:
\begin{verbatim}
mdp_matrix A,B;
A=Random.SU(3);
B=exp(inv(A))*hermitian(A+5);
\end{verbatim}

reads like 
\begin{equation}
\begin{tabular}{l}
$A$ and $B$ are matrices \\ 
$A$ is a random $SU(3)$ matrix \\ 
$B=e^{(A^{-1})}(A+5\cdot \mathbf{1})^H$%
\end{tabular}
\end{equation}

Notice that each matrix can be resized at will and is resized automatically
when a value is assigned.

\section{Lattice, Site and Field}

An \texttt{mdp\_lattice} is a container for \emph{topology} and \emph{%
partitioning} information about the sites. In more abstract terms a lattice
is any collection points (vertices) embedded in a multi-dimensional space
and connected with directional links. The set of links determines the
lattice topology and the boundary conditions. The term partitioning refers
to the function that assigns each site (vertex) to one of the parallel
process. A lattice, by default, is a mesh.

Each lattice is partitioned over the parallel processes at runtime. There is
a default topology and default partitioning but it is possible pass any
topology and partitioning functions to the \texttt{mdp\_lattice} constructor.

A lattice also contains a parallel random number generator: each site of
each lattice has its own independent random number generator.

On each lattice it is possible to allocate one or more fields. Some fields
are built-in, for example: \texttt{mdp\_complex\_field}, \texttt{%
mdp\_vector\_field}, \texttt{mdp\_matrix\_field}, etc. All of them extend
(inherit from) \texttt{mdp\_field\TEXTsymbol{<}mdp\_complex\TEXTsymbol{>}}.

Class \texttt{mdp\_complex} can be used to declare any type of field. For
example:
\begin{verbatim}
class W {
public: int w[10];
};
int L[]={30,30};
mdp_lattice plane(2,L);
mdp_field<W> psi(plane);
\end{verbatim}

declares a $30\times 30$ lattice (\texttt{plane}) and a field (\texttt{psi}%
), that lives on the \texttt{plane}. The field variables of \texttt{psi}, 
\texttt{psi(x)} assuming \texttt{x} is an \texttt{mdp\_site} of \texttt{plane%
}, belong to class \texttt{W}.

Each user-defined field can be saved:
\begin{verbatim}
psi.save("filename");
\end{verbatim}

loaded
\begin{verbatim}
psi.load("filename");
\end{verbatim}

and synchronized
\begin{verbatim}
psi.update();
\end{verbatim}

as any of the built-in fields.

\section{Optimization Issues}

Once an \texttt{mdp\_lattice} object is declared the constructor of class 
\texttt{mdp\_lattice} performs the following operations:

\begin{itemize}
\item[$\bullet$]  Declares a parallel random number generator associated to each site
(it uses the Marsaglia random number generator).

\item[$\bullet$]  Builds tables containing topology and partitioning information that
will be used by the fields to optimize (minimize) communication. Basically
each site determines which other sites are its neighbors and where they are
located (on which parallel process). When a new field is created on the
lattice the field will use these tables to create buffers for the
communications.
\end{itemize}

Once a field object is declared the constructor of class \texttt{mdp\_field}
(or derived field) performs the following operations:

\begin{itemize}
\item[$\bullet$]  Each process allocates memory to store the local sites (i.e. sites
that will be managed by the process itself).

\item[$\bullet$]  Each process loops over every other process and determines if the
other process allocated sites that are neighbors of the local sites (this
information is already stored in the tables maintained by the lattice
object). If this occurs the two processes are said to \emph{overlap}: they
have sites in common that need to be synchronized.

\item[$\bullet$]  Each process allocates buffers to store copies of the sites that are
not local but are neighbors of the local ones and need to be synchronized
with the overlapping processes (in this paper we are assuming only
next-neighbor synchronization but actually {\tt MDP} supports also extended
synchronization such as next-to-next-neighbor and more). Buffers are created
according with some conditions: sites synchronized with the same overlapping
process are stored contiguously in memory so that communication can be
performed in a single \emph{send/receive}. These buffers are created
independently by each field.
\end{itemize}

Notice that two processes may be overlapping in respect to a given lattice
and not overlapping in respect to a different lattice in the same program.

Every time a field changes, for example in a parallel loop such as
\begin{verbatim}
forallsites(x) phi(x)=0;
\end{verbatim}

the program notifies the field that its values have been changed by calling
\begin{verbatim}
phi.update();
\end{verbatim}

The method \texttt{update} performs all required communication to copy site
variables that need to be synchronized between each couple of overlapping
processes. These communications are optimal in the sense that:

\begin{itemize}
\item[$\bullet$]  Each process, at each one time, is involved only in one send and one
receive.

\item[$\bullet$]  Two different processes communicate only if they are overlapping in
respect of the lattice associated to the field.

\item[$\bullet$]  If two different processes are overlapping, they perform a single
send/receive of all sites variables that are synchronized between the two.

\item[$\bullet$]  Only the sending process needs to create a temporary buffer. The
receiving process receives the site variables in the same buffer where they
are normally stored without reordering (and without need for a temporary
buffer).
\end{itemize}

We will refer to our set of communication rules as a ``communication
policy'' (it is possible, in principle, to change this policy to deal with
non-standard network solutions). Although our communication policy does not
overlap communication with computation it has the advantage of minimizing
network jam and calls to send/receive. Hence this communication policy is
almost insensitive to network latency and is dominated by network bandwidth.
Benchmarks are application dependent since parallel efficiency is greatly
affected by the lattice size, by the amount of computation performed per
site, processor speed and type of interconnection. In many typical
applications, like the one described in the preceding example, the drop in
efficiency is less than 10\% up to 8 nodes (processes) and less than 20\% up
to 32 (our tests are usually performed on a cluster of Pentium 4 PCs
(2.2GHz) running Linux and connected by Myrinet).

\section{Conclusions}

{\tt MDP}\ is a powerful and reliable tool for developing efficient parallel
numerical applications. Even if, on the one side, {\tt MDP} is still undergoing
development, on the other side, all of the features here described are fully
functional and have been tested in real-life applications. For example {\tt MDP}
constitutes the core of the {\tt FermiQCD} project \cite{mdp2} developed by the
University of Southampton (UK) and Fermilab (Department of Energy).
{\tt FermiQCD}\ is collection of parallel algorithms for Quantum Chromo Dynamics
computations. The typical {\tt FermiQCD} problem is equivalent to solving
iteratively a system of stochastic differential equations in a 4-dimensional
space. Typical field variables are vectors of complex matrices. {\tt FermiQCD}
programs are used in production runs in parallel on 8 or more nodes.

We believe {\tt MDP}\ could be a useful tool for scientists developing parallel
numerical applications. {\tt MDP} version 2.0 (current) is open source and is free
for research and educational purposes.

\smallskip \textbf{\smallskip Project web pages:}

\begin{itemize}
\item[$\bullet$]  \texttt{http://www.pheonixcollective.org/mdp/mdp.html} (license and
source code)

\item[$\bullet$]  \texttt{http://www.fermiqcd.net} (Lattice QCD\ applications)
\end{itemize}


\begin{thebibliography}{9}
\bibitem{mdp1}  M. Di Pierro, ``Matrix Distributed Processing: ...'',
Computer Physics Communications, \textbf{141} (2001), pp. 98-148
[http://xxx.lanl.gov/abs/hep-lat/0004007]. \emph{Note: this paper describes
version 1.3 of {\tt MDP}. The current version is 2.0}

\bibitem{mdp2}  M. Di Pierro, ``{\tt FermiQCD}'', Nucl. Phys. Proc. Suppl. \textbf{106} (2002) 1034-1036 [http://xxx.lanl.gov/abs/hep-lat/0110116]
\end{thebibliography}
\end{document}